\documentclass[structabstract]{aa}  
\usepackage{graphicx}
\usepackage{txfonts}
\begin{document}
\title{IRAS 18357-0604  - an analogue of the galactic yellow hypergiant IRC +10420?\thanks{Partially based on service observations
made with the {\em William Herschel} Telescope operated on the island of La Palma by the Isaac
Newton Group in the Spanish Observatorio del Roque de los Muchachos of the Instituto
de Astrof\'{i}sica de Canarias.}}

\author{J.~S.~Clark\inst{1}
\and I.~Negueruela\inst{2}
\and C. Gonz\'{a}lez-Fern\'{a}ndez\inst{2,3}}
\institute{
$^1$Department of Physics and Astronomy, The Open 
University, Walton Hall, Milton Keynes, MK7 6AA, United Kingdom\\
$^2$Departamento de F\'{i}sica, Ingenier\'{\i}a de Sistemas y Teor\'{i}a de la Se\~{n}al, Universidad de Alicante, Apdo. 99,
E03080 Alicante, Spain\\
$^3$Institute of Astronomy, University of Cambridge, Madingly Road, Cambridge, CB3 0HA, UK}

   \abstract{Yellow hypergiants represent a short-lived evolutionary episode experienced by massive stars as
they transit to and from a red supergiant  phase. As such,  their properties provide a critical test of stellar
evolutionary theory, while recent observations unexpectedly suggest that  a subset may explode as type II
supernovae.}
{The galactic yellow hypergiant IRC +10420 is a cornerstone system for understanding this phase
 since it is the strongest post-RSG  candidate known, has demonstrated real-time evolution across the Hertzsprung-Russell diagram and been subject to extensive  mass 
loss. In this paper we report on the discovery of a  twin of IRC +10420 - IRAS 18357-0604.}
{Optical and near-IR spectroscopy are used to investigate the physical properties of IRAS 18357-0604 and also 
provide an estimate of its  systemic velocity, while near- to mid-IR photometry probes the nature of its circumstellar environment.}
{These observations reveal pronounced spectral similarities between  IRAS 18357-0604 and IRC +10420,
 suggesting comparable temperatures and wind geometries. IR photometric data reveals a similarly dusty 
circumstellar environment, although historical mass loss appears to have been heavier in IRC +10420. 
The systemic velocity implies a distance  compatible with the red supergiant-dominated complex at the base of the 
Scutum Crux arm; the resultant luminosity determination is  consistent with a physical association but suggests a
lower initial mass than inferred for IRC +10420 ($\lesssim20M_{\odot}$ versus $\sim40M_{\odot}$). 
Evolutionary predictions for the physical properties of supernova progenitors derived from 
 $\sim18-20M_{\odot}$ stars - or  
$\sim12-15M_{\odot}$ stars that have experienced enhanced mass loss as red supergiants -
compare favourably with  those  of IRAS 18357-0604, which in turn appears to be  similar to the
the progenitor of  SN2011dh; it may therefore provide an important insight into the nature of the apparently 
H-depleted yellow hypergiant progenitors of some type IIb SNe.}
{}

\keywords{stars:evolution - stars:early type - stars:individual:IRC +10420}

\maketitle

\section{Introduction}
Yellow hypergiants (YHGs) are thought to represent a short lived episode of the post-main 
sequence (MS) evolution of massive stars; first encountered on a
 redwards passage across  the Hertzsprung-Russell (HR) diagram and, for  more massive stars, on a subsequent 
post-red  supergiant (RSG) loop to higher temperatures. The precise mass range of stars that encounter this
 phase is a sensitive function of a number of stellar parameters such as initial mass and mass loss rate, 
rotational velocity and metalicity (e.g. Ekstr\"{o}m et al. \cite{ekstrom}), making their properties and
 population statistics critical tests of current stellar evolutionary theory. Moreover, with observational 
indications that YHGs drive  extensive mass loss (transient rates $> 10^{-4}M_{\odot}$yr$^{-1}$; 
Castro-Carrizo et al. \cite{castro}, Lobel et al. \cite{lobel}) they may also play an important role in 
mediating the formation of Wolf-Rayet stars by stripping away the H-rich mantle of their progenitors.

Given this, it is unfortunate that  YHGs are amongst the rarest sub-types 
of massive stars known, apparently reflecting the 
relative brevity of this evolutionary phase.  de Jager (\cite{dejager}) and  de Jager \& Nieuwenhuijzen (\cite{dj}) suggested that the most luminous YHGs 
are post-RSG stars on a blue loop across the HR diagram; unfortunately observationally identifying such objects 
 has proved difficult.  Oudmaijer et al. (\cite{oudmaijer09}) list three  post-RSG stars 
- IRC +10420 (see also Jones et al. \cite{jones}), HD179821 and RSGC1-F15 (cf. Davies et al. \cite{davies08}), with  a further
three examples found in the literature - $\rho$ Cas (Lobel et al. \cite{lobel}), HD8752 (Nieuwenhuijzen et al. \cite{nieu})
and IRAS 17163-3907 (Lagadec et al. \cite{lagadec}). Of these, the most compelling case is provided by IRC +10420 due to a N-enriched chemistry, presence of a 
massive ejection nebula and apparent real time bluewards evolution across the HR diagram (Klochkova et al. \cite{kloch97}, Castro-Carrizo et al. 
\cite{castro}, Oudmaijer \cite{oudmaijer98}). Currently demonstrating the 
earliest and richest emission line spectrum of any (candidate) galactic YHG as a result 
of its high luminosity, temperature and mass loss rate, it represents a cornerstone system for understanding the transition of stars away from the 
RSG phase to luminous blue variable (LBV), Wolf-Rayet (WR) or supernova (SN).

As part of an ongoing  $I$-band spectroscopic survey of the environs of the RSG-dominated cluster complex/association at the base of the Scutum-Crux arm (RSG1-5; 
Figer et al. 
\cite{figer}, Davies 
et al. \cite{davies07b},\cite{davies08}, Clark et al. \cite{clark09}, Negueruela et al. \cite{negueruela10}, \cite{negueruela11}, \cite{negueruela12}) we 
observed the bright but poorly studied  IR source IRAS 18357-0604 (henceforth IRAS~1835$-$06; =2MASS J18382341-0601269) located 14' from 
RSGC2/Stephenson 2,  revealing an 
unusually rich emission line spectrum. With the optical region 
inaccessible due to high interstellar reddening, follow-up near-IR observations    were made, which 
confirmed  a striking similarity to IRC +10420; in the remainder of this paper we present and discuss 
these data and their implications.

\begin{figure*}
\includegraphics[width=11cm,height=18cm,angle=270]{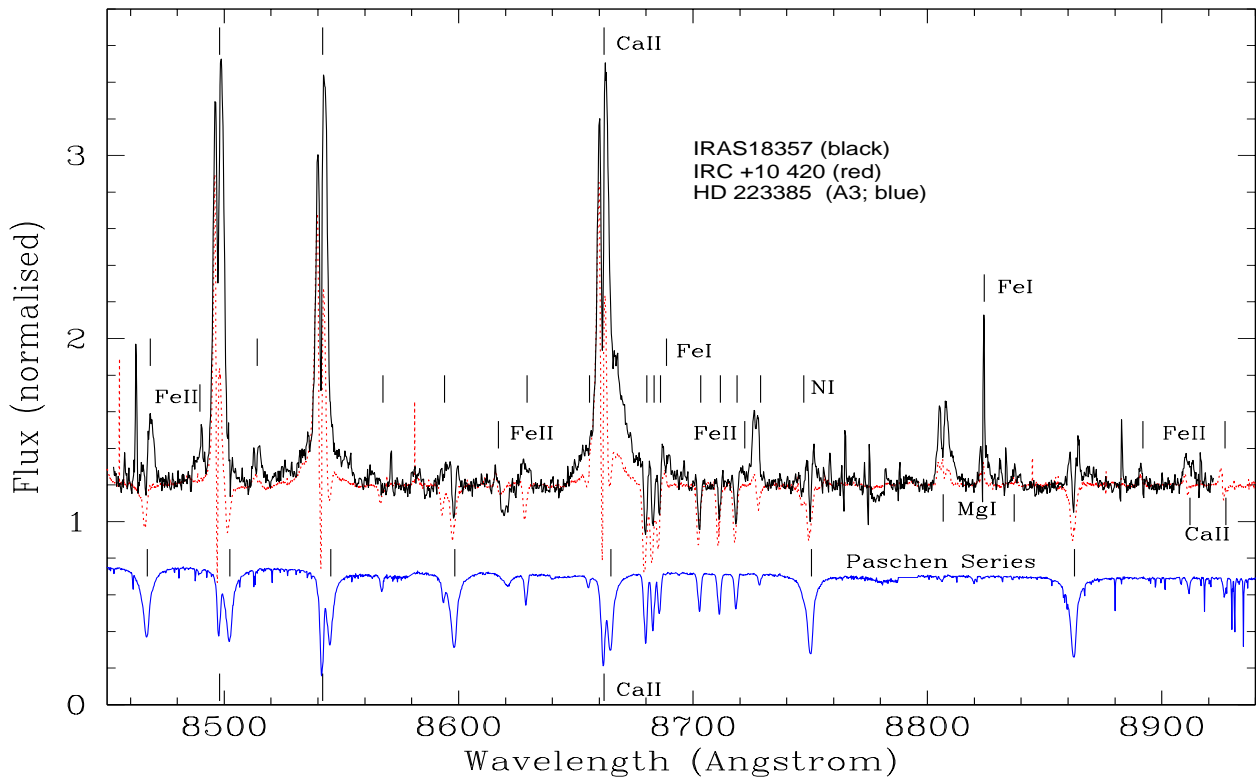}
\caption{I band spectrum of IRAS~1835$-$06 (black, solid lines) plotted against IRC +10420 (red, dotted) and HD223385 (A3 Ia; blue, solid). Spectra have been shifted in wavelength to account for their individual systemic  radial 
velocities as measured from the N\,{\sc i}$\lambda\lambda$ 8703.24,8711.69,8718.82
 photospheric lines. Prominent  emission and 
photospheric lines are indicated.}
\end{figure*}
\section{Data Acquisition \& Reduction}

The $I$-band spectrum was obtained on the night of 2012 July 7 with
the fibre-fed dual-beam AAOmega spectrograph on the 3.9~m
Anglo-Australian Telescope (AAT) at the Australian Astronomical
Observatory. The instrument was operated with
the Two Degree Field ("2dF") multi-object system as front-end. Light
is collected through an optical fibre with a projected diameter of
$2\farcs1$ on the sky and fed into the two arms via a dichroic
beam-splitter with crossover at 5\,700\AA. Each arm of the AAOmega
system is equipped with a 2k$\times$4k E2V CCD detector (the red arm
CCD is a low-fringing type) and an AAO2 CCD controller. Due to the
high reddening to the cluster and short exposure times, only the red
arm registered usable spectra. The red arm was equipped with the
1700D grating, blazed at 10\,000\AA. This grating provides a resolving
power $R=10\,000$ over slightly more than 400\AA. The central
wavelength was set at 8\,600\AA.The exact wavelength range observed
for each spectrum depends on the position of the target in the 2dF
field.

Data reduction was performed using the standard automatic reduction
pipeline {\tt 2dfdr} as provided by the AAT at the time. Wavelength
calibration was achieved with the observation of arc lamp spectra
immediately before each target field. The lamps provide a complex
spectrum of He+CuAr+FeAr+ThAr+CuNe. The arc line lists were revised
and only those lines actually detected were given as input for {\tt
2dfdr}. This resulted in very good wavelength solutions, with rms
always $<0.1$ pixels. Sky subtraction was carried out by means of a mean sky spectrum,
obtained by averaging the spectra of 30 fibres located at known blank
locations. The sky lines in each spectrum are evaluated and used to
scale the mean sky spectrum prior to subtraction.

Near infrared (NIR) spectra for IRAS~1835$-$06 were obtained on 2013 August 13
using LIRIS at the WHT telescope, as part of its service programme.
As the source is very bright in this wavelength regime, we chose a
narrow slit (0.65") that, coupled with medium resolution grisms, yields
an average resolving power around 3000 for the J, H and K bands.
The data were taken using the standard scheme for NIR observations,
nodding the source along the slit in an ABBA pattern so to make it
easy to remove sky emission when reducing the spectra. Said reduction
was carried out in the standard fashion using the 
\texttt{lirisdr} package\footnote{http://www.ing.iac.es/astronomy/instruments/liris/liris$\_$ql.html}. 
To correct for telluric absorption, a B9 star was observed
alongside IRAS 18357-0604. Its spectrum was compared with an ATLAS
model of the same spectral type, making sure that the differences
around the prominent hydrogen absorption lines were low. The normalised
ratio of the observed standard and the model was used then to correct
our spectra. The selected B9 star, HIP091705, was chosen for its
proximity to the target, yet it turned out to be an unresolved binary
with a late type companion. This leaves some residuals when correcting for telluric absorption that show up as small CO emission bands,
particularly for $\lambda>2.3~\mu m$. Consequently we do not discuss this region of the spectrum further.

The resultant spectra are presented in Figs. 1-3.  For line identification in the $I$-band we relied on the
 detailed analysis  of the spectrum of IRC +10420 by Oudmaijer  (\cite{oudmaijer98}) supplemented by 
the 
study of the LBV Wd1-243 in an early-A spectral type phase by Ritchie et al. (\cite{ritchie09}). In the near-IR 
we utilised the line lists for both IRC +10420 (Yamamuro et al. \cite{yamamuro}) and other emission 
line stars 
(Hamann et al. \cite{hamann}, Kelly et al. \cite{kelly}, Clark et al. \cite{clark99} and Filliatre \& Chaty \cite{filliatre}).

\begin{figure}
\includegraphics[width=7cm,angle=270]{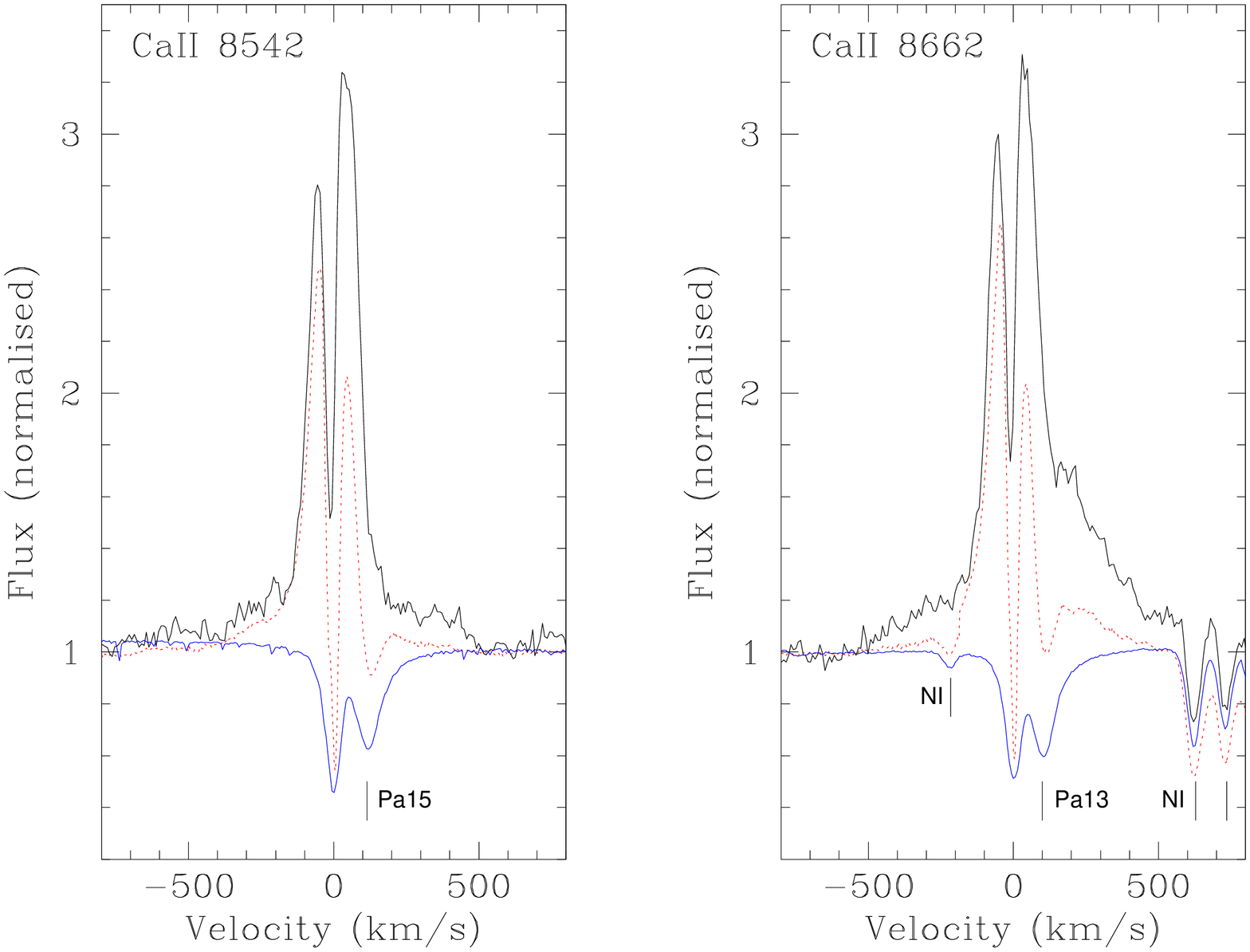}
\includegraphics[width=7cm,angle=270]{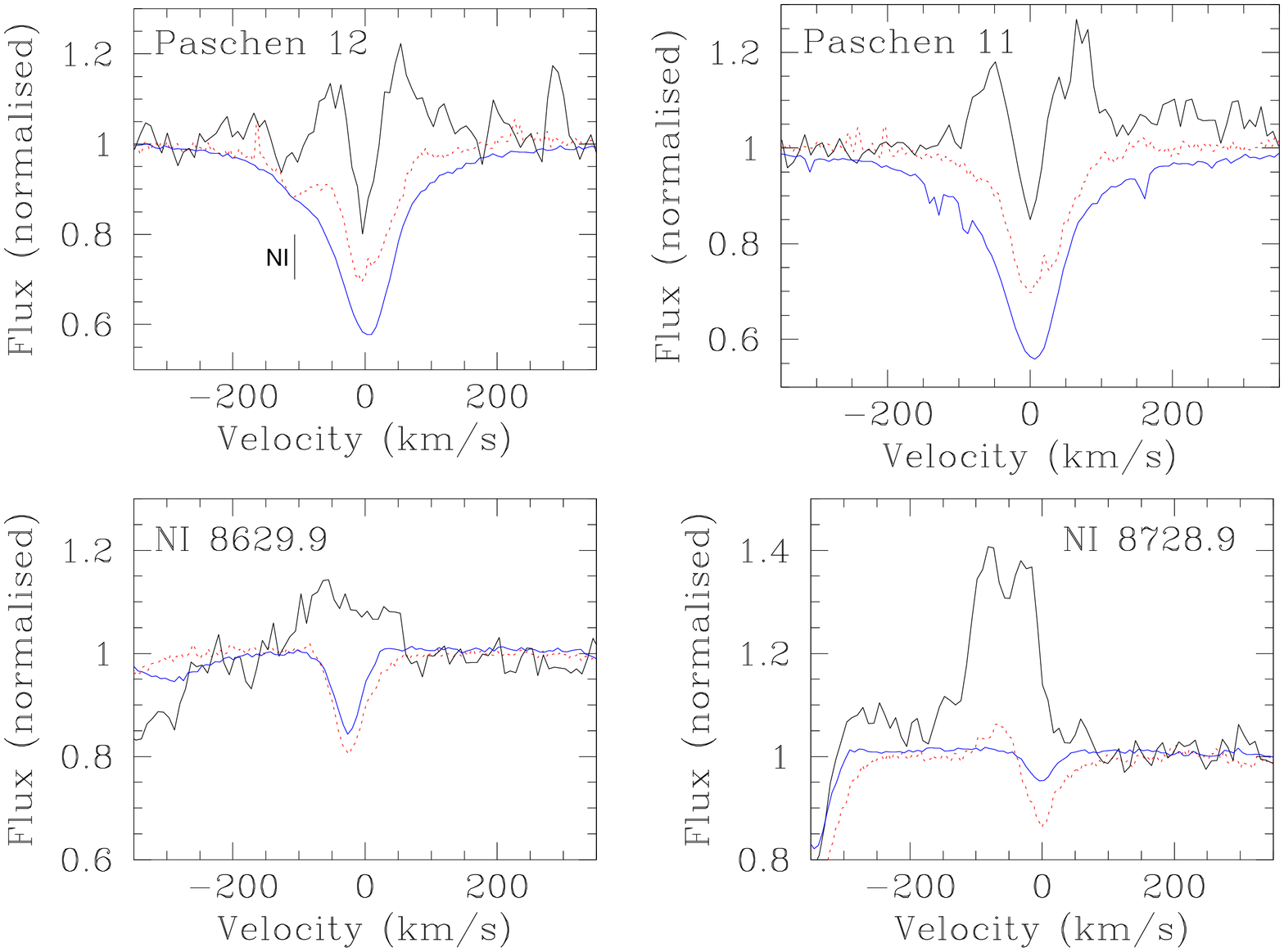}
\caption{Plots of selected $I$-band transitions of the 3 spectra presented in
Fig 1 (IRAS~1835$-$06 - black, solid lines; IRC +10420 - red, dotted; HD223385 - 
blue, solid). Spectra have been shifted in velocity   to account for their individual systemic  radial 
velocities  as described in Fig.~1.}
\end{figure}

\begin{figure}
\includegraphics[width=6.1cm,angle=270]{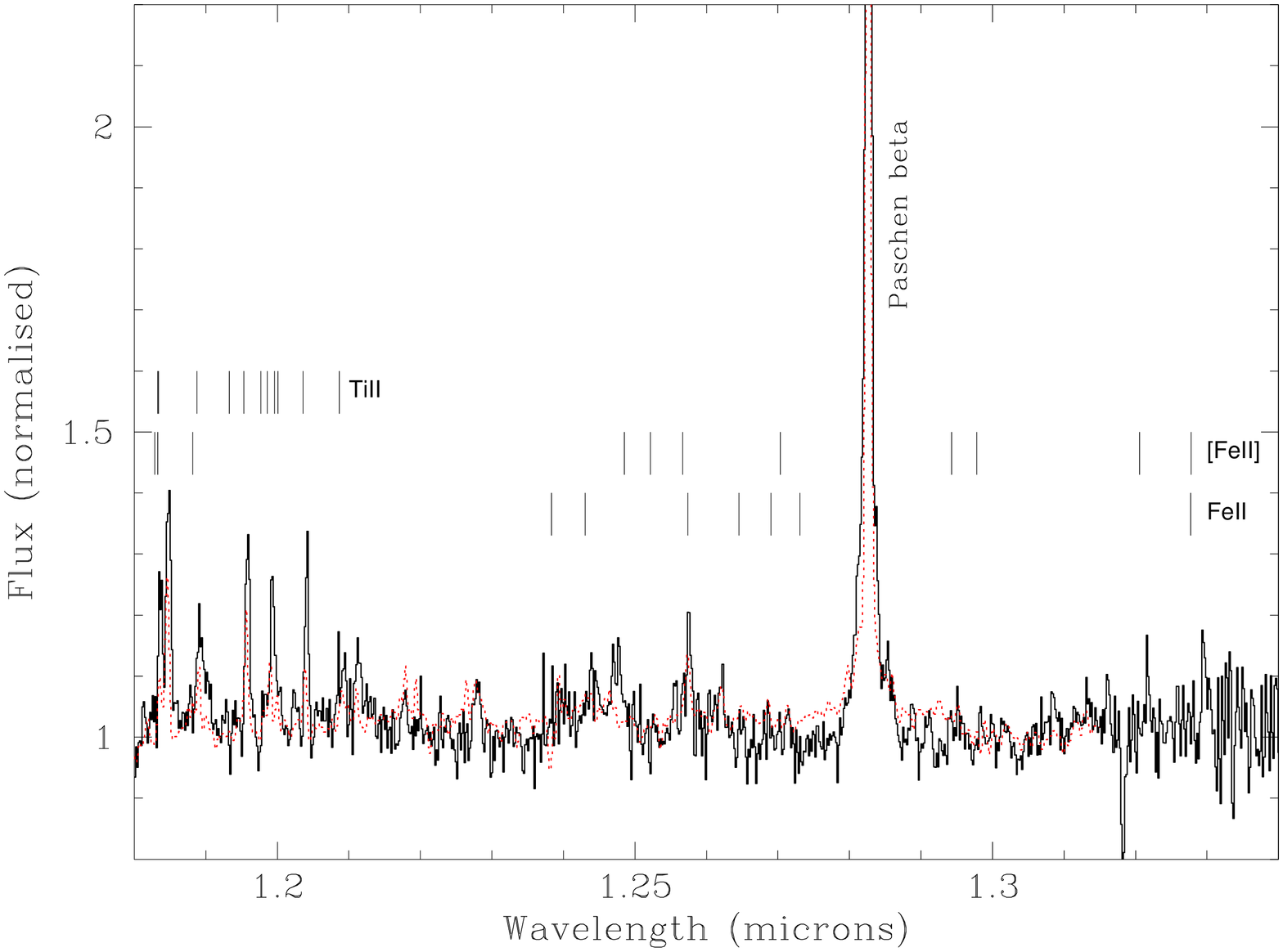}
\includegraphics[width=6.1cm,angle=270]{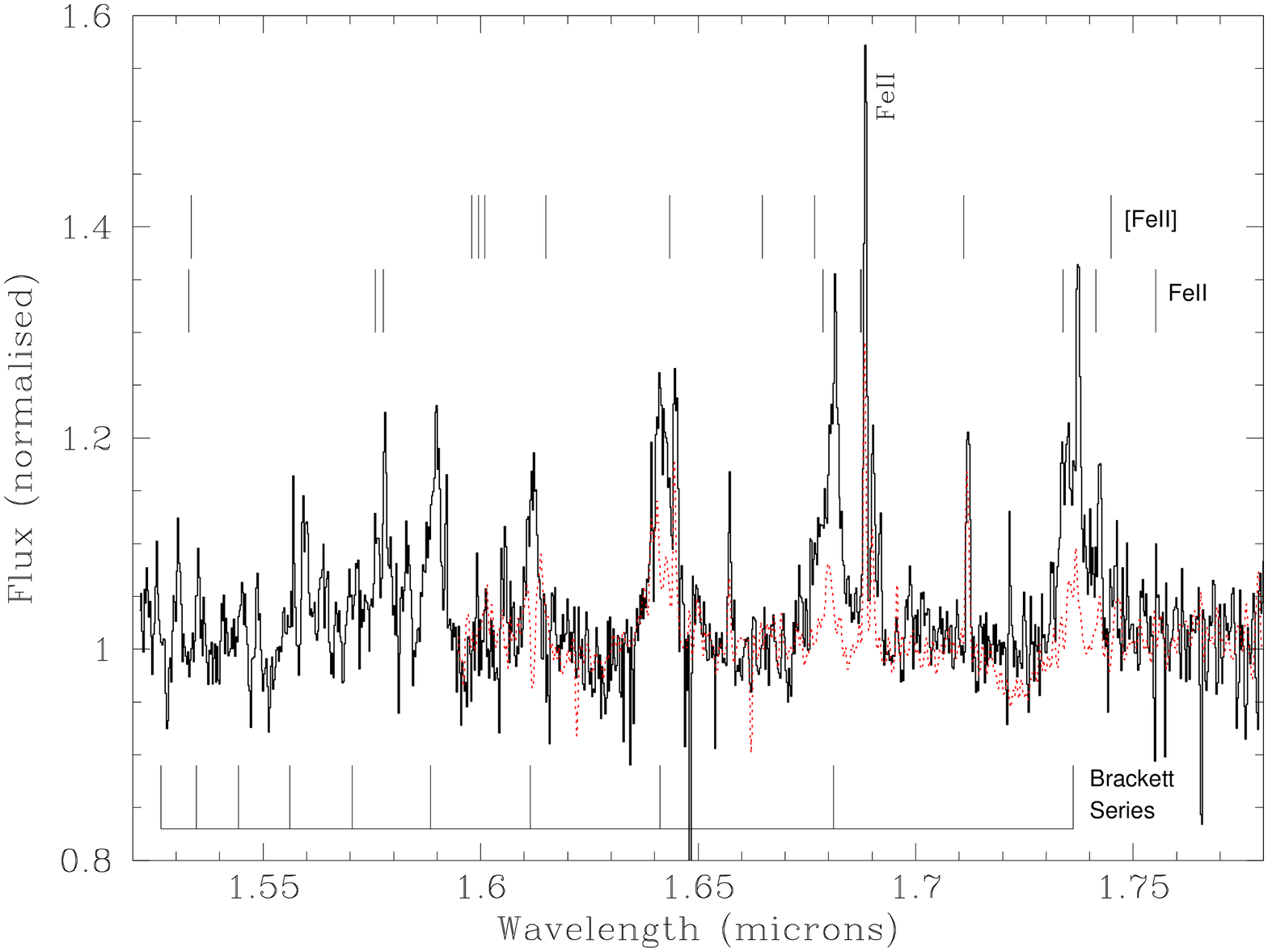}
\includegraphics[width=6.1cm,angle=270]{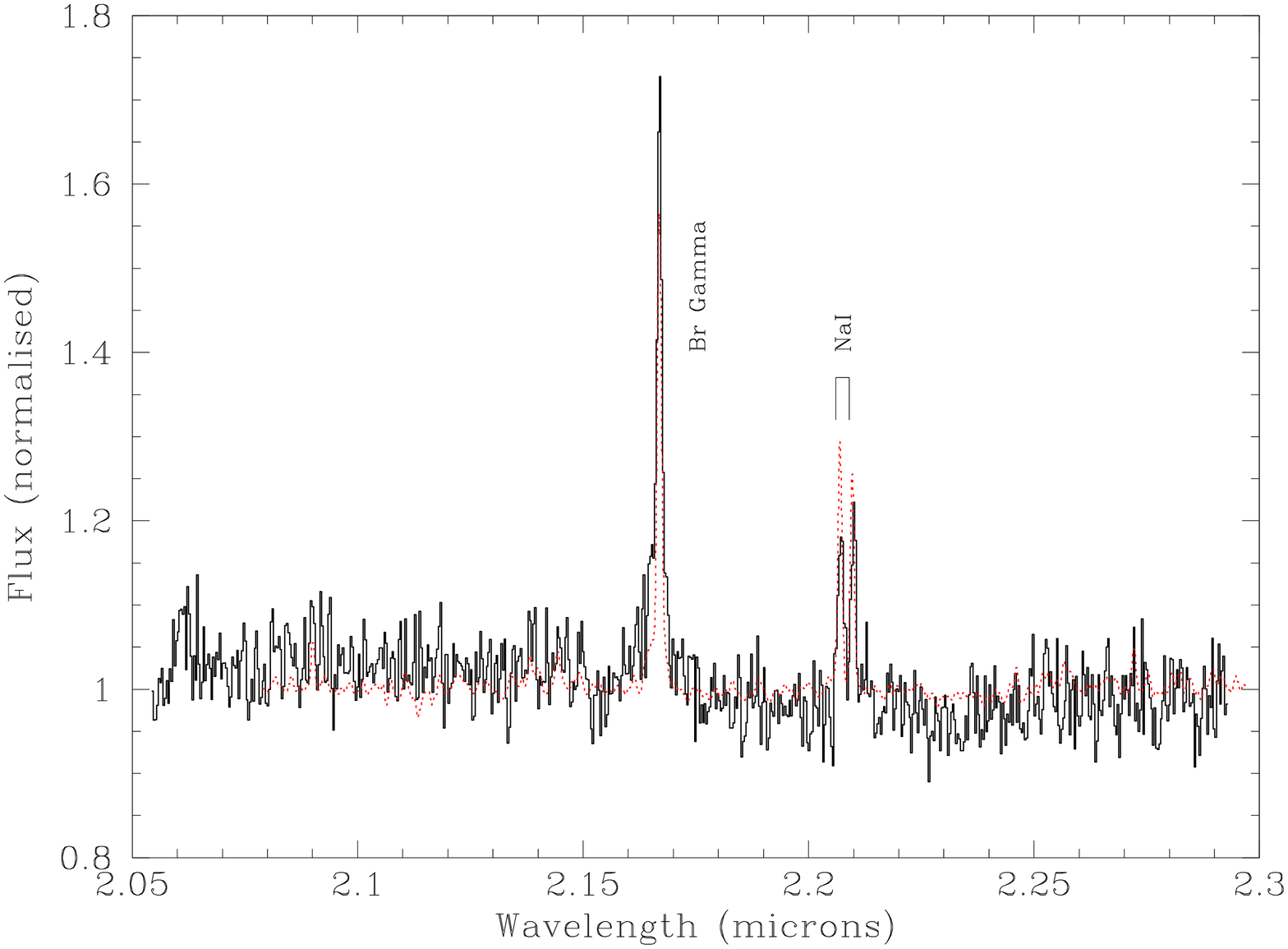}
\caption{$JHK$-band spectra of IRAS~1835$-$06 (black, solid lines) and 
IRC +10420 (red, dotted)
with  transitions due to the Paschen and Brackett series,  Fe\,{\sc 
ii}, [Fe\,{\sc ii}], Na\,{\sc i} and Ti\,{\sc ii} indicated. Peak intensity of the Pa$\beta$ line is 
$\sim$5.2 and $\sim$3.4 times continuum for IRAS~1835$-$06 and IRC +10420 respectively.
Spectra have been corrected for apparent radial velocity as described in Fig. 
1. }
\end{figure}

\section{Results}

As described earlier, the similarities between the $IJHK$-band spectra of IRAS~1835$-$06 and IRC +10420 (Oudmiajer 
\cite{oudmaijer98}, Humphreys et al. \cite{RH02}, Yamamuro et 
al. \cite{yamamuro}; all obtained post transition to spectral type A2 Ia$^+$) are particularly striking (Fig. 1 \& 3). In both cases 
they are dominated by emission 
in the lower transitions of the Paschen and Brackett series as 
well as low excitation metallic species such as Fe\,{\sc 
i}-{\sc ii}, Mg\,{\sc i}, N\,{\sc i}, Na\,{\sc i}, Ti\,{\sc ii} and in particular Ca\,{\sc ii}. None of the  higher excitation lines that are 
present in early, luminous stars (e.g. He\,{\sc 
i}-{\sc ii}, C\,{\sc iii}-{\sc iv} and  N\,{\sc iii}) were seen in emission or absorption. The only 
photospheric absorption lines common to both stars were the  N\,{\sc i} transitions between $\sim8670-8730${\AA}, with  stronger emission in the 
higher Paschen series in IRAS 
1835-06 masking  the photospheric lines present in IRC +10420 (Figs. 1 \& 2; see below). 

Fortuitously, the strength of the N\,{\sc i} absorption lines in the $I-$band  
can function as an approximate temperature diagnostic in late B to early F supergiants, becoming progressively stronger for cooler stars (e.g. Munari 
\& Tomasella \cite{munari97}, Clark et al. \cite{clark05}).
The efficacy of such an approach is shown by comparison of the N\,{\sc i} lines in the spectrum of IRC +10420 to those 
of the A3 Iae (and 
candidate YHG) HD~223385; being marginally stronger and hence indicative of a slightly cooler star in the former. Such a classification is  
entirely consistent with that of early- to mid-A by Oudmaijer et al. (\cite{oudmaijer98}) utilising the  the $\sim4700-4950${\AA} region of the same 
spectrum (and also that of Klochkova et al. \cite{kloch02}).
 Given that these transitions are weaker in IRAS~1835$-$06 one might infer a slightly earlier spectral type for it than 
IRC +10420 ($\sim$A0-2). 
This would also be consistent with the near-IR spectroscopic properties; the lack of He\,{\sc i} 2.112$\mu m$ in absorption suggest $T<10,000$K 
(or A0 or later; Clark et al. \cite{clark11}), while cooler ($T \lesssim 7,500$K; spectral classes F-G) YHGs such as $\rho$ Cas and HR8752 appear unable to support 
the Brackett series in emission (Yamamuro et al. \cite{yamamuro}).

Turning to the emission component of the spectrum and  such  cool temperatures are fully consistent with the presence of both 
Mg\,{\sc i} and strong Ca\,{\sc ii} emission which, given their ionisation potentials (7.65eV and 11.9eV respectively), likely 
arise in a 
neutral H\,{\sc i} region  (cf. Wd1-243; Ritchie et al. \cite{ritchie09}).  While  emission in the Ca\,{\sc ii} lines is stronger in IRAS 
1835-06 than IRC +10420, the profiles in both stars show a remarkable resemblance in both morphology and kinematics; comprising 
high velocity wings  and a narrow, double peaked component (Fig. 2). Following IRC +10420, the 
emission in the high velocity wings of IRAS 
1835-06 increases
from Ca\,{\sc ii} 8498.02{\AA}, through 8542.09{\AA} to 8662.14{\AA}; furthermore the excess emission in the red  over the blue wing of the latter transition
in IRC +10420 is also replicated (Fig. 2). The high velocity emission wings are attributed to Thompson (electon) 
scattering in IRC +10 420 and such a process appears likely here too. 
However, unlike IRC +10420, the red peak is stronger than the blue in all three 
transitions in  IRAS~1835$-$06.
While emission is present in  the (unblended) Pa11 \& 12 lines of IRAS~1835$-$06 (Fig. 2; seen  in 
absorption in IRC +10420), their modest 
strength 
 suggests that  it is unlikely that blending of Pa13, 15 and 16 with their adjacent Ca\,{\sc ii} lines could cause this reversal. 
Where unblended, the Paschen lines demonstrate the  `shell-like' emission profile characteristic of a subset of classical Be stars (Sect. 4). Although hampered by the lower S/N observations, 
where they are discernable the trend to stronger  emission in IRAS~1835$-$06
 is also present in the remaining weak, low excitation metallic transitions, which  
also  appear broader than in IRC +10420,  with several appearing double peaked.  
A particularly notable  feature of the IRC +10420 spectrum is the occurence
 of inverse P Cygni profiles in some of these lines (e.g. N\,{\sc i}$\lambda$8728.9{\AA}; Oudmaijer \cite{oudmaijer98}, see
also Humphreys et al. \cite{RH02}) and a variant of this behaviour  is also
present in IRAS1835-06, where the pure (double peaked) emission component in this 
transition appears similarly blueshifted (Fig. 2). 

Finally, we may estimate the apparent systemic velocity of IRAS~1835$-$06 from the N\,{\sc 
i}$\lambda\lambda 8703.24,8711.69, 8718.82$ lines. Measurement of the systemic velocity shifts in these lines in the spectrum 
of IRC +10420 reveals a mean value of $\sim70$kms$^{-1}$; broadly consistent with that derived from CO rotational 
lines (75kms$^{-1}$; Jones et al. \cite{jones}, Oudmaijer et al. 
\cite{oudmaijer96}) and hence not subject to the anomalous redshift evident in some forbidden lines ($10-20$kms$^{-1}$; Oudmaijer \cite{oudmaijer98}). 
Consequently, we are reassured that the mean value  for IRAS~1835$-$06 ($\sim90\pm3$kms$^{-1}$) reflects the true systemic  velocity, noting 
a  correction for 
transformation to the Local Standard of Reference (LSR) system in this direction of $\sim14$kms$^{-1}$. For comparison H\,{\sc ii} regions in the 
vicinity of Ste2 have $v_{\rm LSR} \sim 90$kms$^{-1}$, while cluster  members have been identified in the 
range $100$kms$^{-1} \leq v_{\rm LSR} \leq 
120$kms$^{-1}$, which includes velocity outliers such as Ste2-17 and Ste2-23 (Davies et al. \cite{davies07b}, 
Negueruela et al. \cite{negueruela12}). We therefore adopt the working hypothesis that IRAS~1835$-$06 is located at the 
same  
distance ($\sim 6$kpc) as Ste2 and associated star forming region for the remainder 
of this paper.

\section{Discussion}
\subsection{Spectral comparisons}

Given that YHGs, (cool phase) LBVs and supergiant B[e]/A[e]  stars are co-located on the HR 
diagram and show similar spectral morphologies (dominated by emission in H\,{\sc i} and low excitation metals),
 it is worth investigating whether IRAS~1835$-$06 fulfills the classification criteria of either of the latter two classes of star. $I$-band 
spectroscopy of  LBVs shows a diversity of behaviour. During their cool-phases HR Car and  AG Car
show pronounced differences with respect to IRAS~1835$-$06; the spectrum of HR Car  being  characterised by weak P Cygni emission in the Paschen  series, which are stronger and supplemented by N\,{\sc 
i} emission in AG Car -  in neither case is Ca\,{\sc ii} emission 
prominent (Machado et al. \cite{machado}; Jose Groh, priv. comm. 2013). 
Conversely, in systems where it is 
prominent, neither the Paschen series nor low excitation metal lines are in emission (e.g. S Dor \& Wd1-243; 
Munari et al. \cite{munari97}, Clark \& Negueruela \cite{clark04}).

Only two A[e] supergiants are known and only one   - 3 Pup -  has published $I-$band data. Ca\,{\sc ii} emission is 
present, although the line profiles are narrower than in IRAS~1835$-$06 and also lack a 
deep central absorption trough and strong emission wings (Chentsov et al. \cite{chentsov}). These  differences are also present in these 
transitions in hotter B[e] supergiants (Aret et al. \cite{aret}). No systemic surveys of either LBVs or B[e]/A[e] supergiants are
available in the $JH$-bands, but in the $K$-band both IRC +10420 and IRAS~1835$-$06 lack the weak Fe\,{\sc ii} and 
Mg\,{\sc ii} emission of the coolest LBVs (Clark et al. \cite{clark11}), but demonstrate stronger Na\,{\sc 
i} emission with respect to Br$\gamma$ than seen in B[e] stars (Oksala et al. \cite{oksala}).

 We conclude that  IRAS~1835$-$06 most closely resembles the YHG IRC +10420 in comparison to both LBVs and 
B[e]/A[e] supergiants, although we caution that these classifications are likely not mutually 
exclusive. For example the sgB[e] star S18  demonstrates LBV-like photometric and spectroscopic 
variability (Clark et al. \cite{clark13}). Likewise the cool-phase spectrum of the LBVs R71 and R143 (Munari et al. \cite{munari}, Mehner et al. \cite{mehner}) are 
almost indistinguishable from those of the YHGs Wd1-12 and 16a (Clark et al. \cite{clark05}); in the absence of 
variability both R71 and R143 would have also been classified as YHGs. The contrast between the $I$-band spectra of  IRC +10420 and 
IRAS~1835$-$06 (dominated by emission) and the YHGs within Westerlund~1 (dominated by absorption)  also highlights an
additional issue: the heterogeneous nature of YHG spectra, even for stars of similar temperature such as these four.  
Potential explanations for this diversity are the wide range of  luminosities spanned by known examples ($(L_{\rm bol}/L_{\odot})\sim 
5.4-6.2$; Davies et al. \cite{davies08}, Clark et al. in prep.)\footnote{Although in this case the luminosities of IRC 
+10420, Wd1-12 and 16a 
are expected to be broadly comparable (e.g. Oudmaijer et al. \cite{oudmaijer09}).} and presence of both   pre- and post-RSG examples, with 
the properties of the latter expected to be 
influenced by the chemical enirchment and substantial mass loss that they will have 
 experienced.

The unusual emission line profiles  that characterise the IRC +10420 spectrum have led to significant debate regarding 
the  geometry of the circumstellar environment/wind (c.f Jones et al. \cite{jones}, Oudmaijer et al. 
\cite{oudmaijer96}, Humphreys et al. \cite{RH02}, Davies et al. \cite{davies07a}).
 Recent interferometric observations suggested a mass loss rate of $\sim1.5-2\times10^{-5}M_{\odot}$yr$^{-1}$
and  confirmed asphericity in the inner wind (Driebe et al. \cite{driebe}),  who suggested that either the wind was 
intrinsically asymmetric, with a greater mass loss from the hemisphere pointed towards us or that the wind is spherical, with emission from the receding component blocked by an equatorial disc. Subsequently Tiffany et al. (\cite{tiffany}) suggested
 that IRC +10420 is observed in the  pole-on orientation required by the latter hypothesis, with Oudmaijer \& de Wit (\cite{oudmaijer13})  favouring a geometrical model in which the receding  portion of 
a polar wind  is occulted by the stellar disc, leading to the asymmetric H$\alpha$ and Ca\,{\sc ii} lines.

It would therefore be natural to infer a similarly powerful, asymmetric wind for IRAS~1835$-$06, given its overal spectral similarity. However, 
 the observed blue- to red-peak ratio of the Ca\,{\sc ii} lines in IRAS~1835$-$06 could not arise through obscuration/inclination effects
(which should only act to reduce the strength of the red-shifted peak relative to blue) and so would have to reflect a real asymmetry in wind strengths between different 
hemispheres. Motivated by the shell-like lines in the Paschen series, an alternative explanation might be that both Ca\,{\sc ii} and Paschen series lines arise in an 
equatorial disc viewed `edge-on' (cf. $o$ And; Clark et al. \cite{clark03}). However, classical Be stars observed in such an orientation also demonstrate  shell profiles in 
low-excitation metallic species such as Fe\,{\sc ii}; such behaviour is not reproduced in IRAS~1835$-$06. 

\subsection{IR properties}
 IRAS~1835$-$06 is a highly reddened system ($J=8.54\pm0.02, H=6.28\pm0.02$ and $K=4.63\pm0.02$) for which  no optical photometry is currently  available. 
Adopting the 
intrinsic colour for an $\sim$A2 Ia star ($(J-K)\sim0.11$; Koornneef \cite{koornneef}) leads to $E(J-K)\sim3.8$; with an IR reddening-free parameter $Q_{\rm IR} = 
(J-H)-1.8\times(H-K)\sim-0.7$ (Negueruela et al. \cite{negueruela12}), suggesting a combination of both interstellar extinction and circumstellar emission. Such a 
conclusion is supported by its mid-IR luminosity; IRAS~1835$-$06 is saturated in GLIMPSE/{\em Spitzer} data (Benjamin et al. \cite{benjamin}) 
but is a bright point source 
apparently 
associated with no extended emission in IRAS (IRAS PSC  \cite{IRAS}), {\em Midcourse Source Experiment} (Egan et al. \cite{egan}) 
and WISE (Wright et al. \cite{wright}) 
 data\footnote{WISE fluxes of $F_{3.35{\mu}m}\sim15.7^{+2.0}_{-1.8}$Jy, $F_{4.60{\mu}m}\sim38.3^{+4.7}_{-4.2}$Jy, 
$F_{11.56{\mu}m}\sim19.7^{+1.0}_{-1.0}$Jy and $F_{22.09{\mu}m}\sim18.5^{+0.3}_{-0.3}$Jy,
MSX  fluxes of $F_{8.28{\mu}m}\sim27.3\pm1.1$Jy, $F_{12.13{\mu}m}\sim27.7\pm1.4$Jy, $F_{14.7{\mu}m}\sim24.5\pm1.5$Jy and 
$F_{21.34{\mu}m}\sim18.0\pm1.1$Jy and IRAS fluxes of $F_{12{\mu}m}\sim27$Jy, $F_{25{\mu}m}\sim19.1$Jy
and $F_{60{\mu}m}\sim17.1$Jy.}. Likwise Kwok et al. (\cite{kwok}) report  a featureless low resolution IRAS spectrum  that is flatter than expected for a stellar source. 

Potential  contamination of the  near-IR photometry by dust emission renders a direct reddening determination unreliable.
 Consequently, given its proximity to Ste2 (Sect. 3) we make use of the mean reddening towards the cluster ($A_{K}\sim1.0\pm0.25$; Froebrich \& Scholz \cite{froebrich}). After deredening these data (following Messineo et al. \cite{messineo})
we may then  determine the near/mid-IR 
colour  indices employed by Davies et al. (\cite{davies07b})\footnote{$(K-A)\sim3.3$, $(K-C)\sim4.2$ and 
$(K-D)\sim3.6$}, from which we find that IRAS~1835$-$06 supports a larger IR excess than  any cluster member - including the post-RSG candidate 
Ste2-49 -  confirming the presence of warm circumstellar dust (Fig. 4). Given the spectral similarities of  IRAS~1835$-$06 and  IRC 
+10420, if we assume both support comparable winds  (Driebe et al. \cite{driebe}) then following the analysis of Kochanek 
(\cite{kochanek}) dust should not currently be able to condense in either system. This would imply that IRAS~1835$-$06 has recently undergone an 
episode of enhanced mass loss and it is tempting to suggest that it has just exited a RSG-phase during which the circumstellar dust formed.
However, it appears to lack the massive, spatially resolved ejecta that characterises both IRC +10420 and IRAS 17163-3907 ($1M_{\odot}$ and $4M_{\odot}$ respectively), although these are by no means ubiquitous amongst YHGs, being absent for $\rho$ Cas and HR 8752  (Humphreys et al. \cite{RH97}, Schuster et al. \cite{schuster}, Castro-Carrizo et al. \cite{castro},  Lagadec et al. \cite{lagadec}).

Finally, using the mean cluster reddening and the $J$-band magnitude - since it is least likely to be contaminated 
by dust emission - we may estimate the current bolometric luminosity of IRAS~1835$-$06. Following Okumura et al. 
(\cite{okumura}) we may convert $A_K$ to $A_J$ and then for a distance modulus of 13.9 ($\sim6$kpc) we find 
$M_J\sim-8.3\pm0.7$. Then employing the colour relationships from Koornneef (\cite{koornneef}) and the relevant bolometric 
correction for early A supergiants (Clark et al. \cite{clark05}) we arrive at log$(L_{\rm bol}/L_{\odot}) \sim5.2{\pm}0.3$, where the 
error quoted is dominated by uncertainties in the interstellar reddening.
This value suggests that IRAS~1835$-$06 is intrinsically less luminous than IRC +10420
 (log$(L_{\rm bol}/L_{\odot}) \sim5.7-5.8$; e.g. Jones et al. \cite{jones}, Oudmaijer et al. \cite{oudmaijer96}) and hence also 
less massive ($\lesssim20M_{\odot}$ versus 
$\sim40M_{\odot}$), noting that any contribution to the $J$-band flux by warm dust would lead to a corresponding reduction in 
 luminosity.

\begin{figure}
\includegraphics[width=8cm,angle=0]{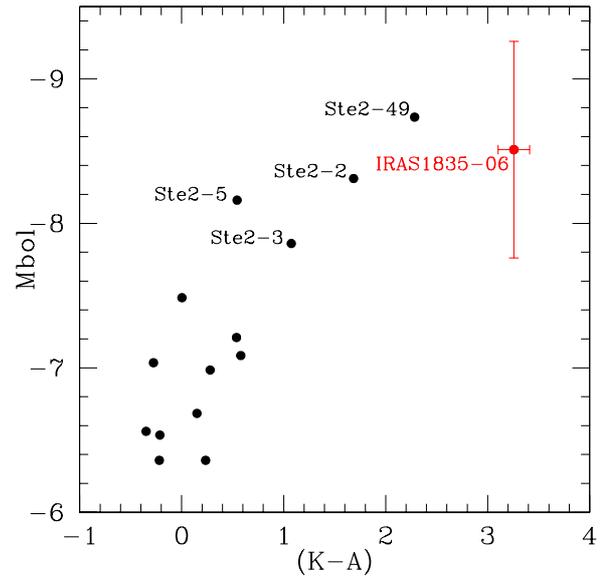}
\caption{Plot of the $(K-A)$ near- to mid-IR colour indice versus 
$M_{\rm bol}$ for IRAS~1835$-$06 and members of Ste2 (following Davies et al. 
\cite{davies07b}). Errors in $M_{\rm bol}$ for IRAS1835-06 are dominated by 
uncertainty in reddening and the value is for an assumed distance of 
$\sim$6kpc. Individual errors for Ste2 members are not presented by Davies et 
al. (\cite{davies07b}, their Fig. 10) but appear to be $\lesssim0.25$mag in $M_{\rm bol}$
 and of order the symbol size in $(K-A)$. In the absence of an IR excess we would 
expect a star to reside at $(K-A)\sim0$.}
\end{figure}

\begin{figure}
\includegraphics[width=8cm,angle=0]{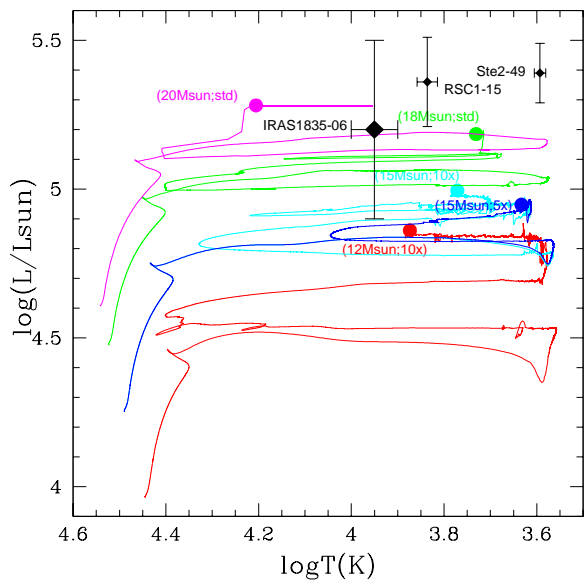}
\caption{HR diagram showing the position of IRAS~1835$-$06 and the post-RSG candidates RSGC1-15 and Ste2-49 (Davies et al. 
\cite{davies07b},\cite{davies08})
assuming a distance of $\sim6$kpc. We conservatively adopt $8,000K < T < 10,000K$ for IRAS~1835$-$06 (appropriate for an A hypergiant).
We also plot
the evolutionary tracks of Georgy (\cite{georgy}) and Groh et al. (\cite{groh13}) with the positions of the resultant SNe indicated by coloured dots. The $18M_{\odot}$ and   $20M_{\odot}$ tracks (green  and purple
 respectively) assume the 
standard mass loss rate prescription of Ekstr\"{o}m et al. (\cite{ekstrom}), the 
$15M_{\odot}$ tracks  $5\times$ and $10\times$ standard (dark and light blue respectively) and the 
$12M_{\odot}$ track assumes a mass loss rate  $10\times$ standard (red).}
\end{figure}

\subsection{Environmental and evolutionary context}
Are the properties we adopt for IRAS~1835$-$06 consistent with a physical association with either Ste2 or the larger
 RSG agglomerate at the base of the Scutum-Crux arm (Sect. 1)? RSGC1 and Ste2  appears to be the youngest and oldest 
clusters within this  complex, suggesting star formation peaking over the last 10-20Myr resulting in
$M_{\rm init}\sim12-22M_{\odot}$ for evolved stars  within this region (Davies et al. 
\cite{davies08})\footnote{An age of $12\pm2$Myr and $17\pm3$Myr for RSGC1 and Ste2 respectively, implying initial masses of 
$18^{+4}_{-2}M_{\odot}$ and $14\pm2M_{\odot}$ for the RSG cohorts.}. If this age range is applicable to the wider RSG association
 then the current luminosity and temperature of IRAS~1835$-$06 are consistent with membership, with 
theoretical models  for rotating stars with $M_{\rm init}\sim18-20M_{\odot}$ reproducing its current  parameters at an age 
comparable to that of RSGC1 (Fig. 5; Ekstr\"{o}m et al. \cite{ekstrom}, Groh et al. \cite{groh13}).  This in turn 
would imply that  IRAS~1835$-$06 could either be the {\em immediate} progenitor of a 
core-collapse type IIb SN, or may instead evolve through an LBV phase prior to encountering this fate, 
 with the $M_{\rm init}\sim18M_{\odot}$ model terminating as a YHG ($T\sim5400$K) and the $M_{\rm init}\sim20M_{\odot}$ terminating as a LBV 
($T\sim19500$K), in concert bracketing the current properties of IRAS~1835$-$06.

More particularly,  are the properties of IRAS~1835$-$06 consistent with an origin within, and subsequent ejection from, 
the older cluster Ste2?\footnote{Assuming a transverse `runaway' velocity of 10kms$^{-1}$  (compared to a 
1$\sigma$ velocity dispersion of $\sim$3kms$^{-1}$ for Ste2; Davies et al. \cite{davies07b}) would place 
IRAS~1835$-$06 14' distant after only 730,000yr; well within the expected lifetime of such a star.}
Stars of such an age are not expected to evolve past the  RSG phase and the current  
luminosity and temperature  of IRAS~1835$-$06 are not consistent with a pre-RSG YHG phase (Ekstr\"{o}m et al. 
\cite{ekstrom}). However Davies et al. (\cite{davies07b}) identify  Ste2-49 as  a putative post-RSG object, 
while Ste2-2, 3 and 5 also have luminosities in excess of  log$(L_{\rm bol}/L_{\odot})\sim5.0$; hence 
comparable to IRAS~1835$-$06. Moreover, all four of these stars  support the most extreme  mass loss 
rates of any  of the members of Ste-2 (as measured from their dust-driven mid-IR excesses; Fig 4) and hence it 
is tempting to posit that IRAS~1835$-$06 is a more evolved descendent of such stars.  

Georgy (\cite{georgy}; Fig. 5) shows that an increase in the mass loss rates adopted for
 RSGs by a factor of 5-10 would enable stars in this mass range to evolve through to a YHG phase 
immediately prior to a SN endpoint, again with properties broadly comparable with those of IRAS~1835$-$06 (cf. their 
$M_{\rm init}\sim12(15)M_{\odot}$ models with a mass loss rate 10$\times$ standard, 
which at SN have log$(L_{\rm bol}/L_{\odot})\sim4.87(5.0)$ and $T_{\rm eff}\sim7400(6000)K$).
 Such stars would be expected to approach the Eddington limit and hence support strong current mass loss rates, implying 
that observationally they should resemble IRAS~1835$-$06 (Jose Groh, priv. comm. 2013).
 
Consequently, if the association of IRAS~1835$-$06 with Ste2 were confirmed it would 
 validate the suggestion that (a subset of) RSGs experience  extreme mass loss rates  and as a result  provide an unique 
insight into the properties  of low luminosity yellow super-/hypergiant SNe progenitors such as that of the type IIb 
event  SN2011dh  (log$(L_{\rm bol}/L_{\odot}) \sim4.9\pm0.2$, $T_{\rm eff}\sim6000\pm200K$ and 
$M_{\rm init}\sim13\pm3M_{\odot}$; Maund et al. \cite{maund}, van Dyk et al. \cite{van}). 
 Indeed, despite their relatively low luminosities, such SNe progenitors would better be classified as hypergiants 
rather than supergiants (as currently described in the literature), given their extreme post-RSG mass loss rates which would  result in rich emission line spectra (cf. Groh et al. \cite{groh13}).

\section{Conclusions}
In spectroscopic terms IRAS~1835$-$06  is a near-twin of the hitherto unique post-RSG YHG IRC +10420. Similarly
demonstrating asymmetric, twin-peaked emission line profiles - indicative of 
aspherical, outflowing material -
IRAS1835-06 reveals   that the complex circumstellar  environment of IRC +10420 is not a pathological case,
 but instead potentially represents a template for this phase of stellar evolution. The mid-IR properties of IRAS~1835$-$06
reveal the presence of a substantial dusty circumstellar component which, although apparently 
not as massive as that of IRC +10420, is fully consistent with identification as a YHG. Given that it appears unlikely that its
 current wind could support dust condensation, this is suggestive of a recent exit from  a high mass loss rate episode (possibly associated with the preceding RSG phase).

A distance of $\sim$6kpc is inferred from the systemic velocity of IRAS~1835$-$06, which  would place it in the RSG `association'
at the base of the Scutum-Crux arm. Such a distance implies log$(L_{\rm bol}/L_{\odot})\sim5.2\pm0.3$; entirely  consistent with 
an origin in the same burst of star formation that yielded the RSG population (although we caution that such a value must be 
regarded as {\em provisional} pending  a re-determination utilising deep optical observations that are free from 
contamination from circumstellar dust).  Likewise its apparent rarity - the only known 
example in comparison to $\sim10^2$ RSGs within the complex - is predicted by current evolutionary models (e.g. Ekstr\"{o}m et 
al. \cite{ekstrom}, Davies et al.  \cite{davies09}). Located 14' from the massive cluster Ste2, identification as a runaway cannot 
be excluded, but an unexpectedly extreme mass-loss rate during the preceding RSG phase would be required to replicate its current properties in such a scenario.

 Nevertheless, either scenario would imply that IRAS~1835$-$06 could represent the {\em immediate} progenitor of a 
core-collapse  SN (Georgy \cite{georgy}, Groh et al. \cite{groh13}) and its current physical properties are indeed similar to those of the YHG progenitor of SN2011dh. IRAS~1835$-$06 may therefore 
shed light on  both the rapid, post-RSG evolution of massive stars  and also the nature of the YHG progenitors  
of some type IIP and IIb SNe (e.g. SN1993J, SN2008cn and SN2009kr; Maund  et al. \cite{maund}), which are thought to arise from 
moderately massive stars which have been progressively stripped of their hydrogen mantle.

\begin{acknowledgements}
This research is partially supported by the
    Spanish Ministerio de Econom\'{\i}a y Competitividad (Mineco)
    under grants AYA2010-21697-C05-05 and AYA2012-39364-C02-02.
The AAT observations have been supported by the OPTICON project
(observing proposal 2012/A015) , which is funded by the European
Commission under the Seventh Framework Programme (FP7). R. Dorda contributed to
this observation and carried out the reduction of the resultant AAOmega 
spectrum. We thanks Cyril Georgy for supplying the evolutionary tracks used 
in the construction of Fig. 5.

\end{acknowledgements}

{}

\begin{thebibliography}{}

\bibitem[2012]{aret}
Aret, A., Kraus, M., Muratore, M. F. \&  Borges Fernandes, M. 2012,
MNRAS, 423, 284

\bibitem[2003]{benjamin}
Benjamin, R. A., Churchwell, E., Babler, B. L., et al. 2003, PASP, 115, 953

\bibitem[2007]{castro}
Castro-Carrizo, A., Quintana-Lacaci, G. Bujarrabal, V., Neri, R. 
\& Alcolea, J. 2007, A\&A, 465, 457

\bibitem[2010]{chentsov}
Chentsov, E. L., Klochkova, V. G. \& Miroshnichenko, A. S. 2010, AtBu, 65, 150

\bibitem[1999]{clark99}
Clark, J. S., Steele, I. A., Fender, R. P. \& Coe, M.J. 1999, A\&A, 348, 
888

\bibitem[2004]{clark04}
Clark, J. S. \& Negueruela, I. 2004, A\&A, 413, L15

\bibitem[2003]{clark03}
Clark, J. S., Tarasov, A. E., Panko, E. A., 2003, A\&A, 403, 239

\bibitem[2005]{clark05}
Clark, J. S., Negueruela, I., Crowther, P. A. \& Goodwin, S. P. 
2005, A\&A, 434, 949

\bibitem[2009]{clark09}
Clark, J. S., Negueruela, I., Davies, B., et al. 2009, A\&A, 507, 1555

\bibitem[2011]{clark11}
Clark, J. S., Arkharov, A., Larionov V., et al. 2011, BSRSL, 80, 361   

\bibitem[2013]{clark13}
Clark, J. S., Bartlett, E. S., Coe, M. J. et al. 2013, A\&A, in press [arXiv1305.0459]

\bibitem[1994]{danks}
Danks, A. C. \& Dennefeld, M. 1994, PASP, 106, 382


\bibitem[2007a]{davies07b}
Davies, B., Figer, D. F., Kudritzki, R.-P. et al. 2007a, ApJ, 671, 781


\bibitem[2007b]{davies07a}
Davies, B., Oudmaijer, R. D. \& Sahu, K. C. 2007b, ApJ, 671, 2059

\bibitem[2008]{davies08}
Davies, B., Figer, D. F., Law, C. J. et al. 2008, ApJ, 676, 1016

\bibitem[2009]{davies09}
Davies, B., Figer, D. F., Kudritzki, R.-P.,et al. 2009, ApJ, 707, 844

\bibitem[1998]{dejager}
de Jager, C. 1998, A\&ARv, 8, 145

\bibitem[1997]{dj}
de Jager, C. \& Nieuwenhuijzen, H. 1997, MNRAS, 290, L50

\bibitem[2009]{driebe}
Driebe, T., Groh, J. H., Hofmann, K.-H. et al. 2009, A\&A, 507, 301

\bibitem[2001]{egan}
Egan, M. P., Price, S. D. \& Gugliotti, G. M. 2001, BAAS, 34, 561

\bibitem[2012]{ekstrom}
Ekstr\"{o}m, S., Georgy, C., Eggenberger et al. 2012, A\&A, 537, 146

\bibitem[2006]{figer}
Figer, D. F., MacKenty, J. W., Robberto, M., et al. 2006, ApJ, 643, 1166

\bibitem[2004]{filliatre}
Filliatre, P. \& Chaty, S. 2004, ApJ, 616, 469

\bibitem[2013]{froebrich}
Froebrich, D. \& Scholz, A. 2013, MNRAS, in press [arXiv:1308.6436]

\bibitem[2012]{georgy}
Georgy, C. 2012, A\&A, 538, L8

\bibitem[2013a]{letter}
Groh, J. H., Meynet, G. \& Ekstr\"{o}m, S. 2013a, A\&A, 550, L7 

\bibitem[2013b]{groh13}
Groh, J. H., Meynet, G., Georgy, C. \& Ekstr\"{o}m 2013b A\&A, 558, A131

\bibitem[1994]{hamann}
Hamann, F., Depoy, D. L., Johansson, S \& Elias, J. 1994 ApJ, 422, 626

\bibitem[1997]{RH97}
Humphreys, R. M., Smith, N., Davidson, K. et al. 1997, AJ, 114, 2778

\bibitem[2002]{RH02}
Humphreys, R. M., Davidson, K. \& Smith, N. 2002, AJ, 124, 1026

\bibitem[1985]{IRAS}
IRAS Point Source Catalogue:1985, US Government Publication Office

\bibitem[1993]{jones}
Jones, T. J., Humphreys, R. M., Gehrz, R. D. et al. 1993, ApJ, 411, 323

\bibitem[1994]{kelly}
Kelly, D. M., Rieke, G. H. \& Campbell, B. 1994, ApJ, 425, 231

\bibitem[1997]{kloch97}
Klochkova, V G.,  Chentsov, E. L. \& Panchuk, V. E. 1997, MNRAS, 292, 19

\bibitem[2002]{kloch02}
Klochkova, V. G., Yushkin, M. V., Chentsov, E. L. \& Panchuk, V. E. 2002, ARep, 46, 139

\bibitem[2011]{kochanek}
Kochanek, C. S. 2011, ApJ, 743, 73

\bibitem[1983]{koornneef}
Koornneef, J. 1983, A\&A, 128, 84

\bibitem[2008]{kraus}
Kraus, M., Borges Fernandes, M., Kub\'{a}t, J. \& de Ara\'{u}jo, F. X. 
2008, A\&A, 487, 697

\bibitem[1997]{kwok}
Kwok, S., Volk, K., Bidelman, W. P., 1997, ApJS, 112, 557

\bibitem[2011]{lagadec}
Lagadec, E., Zijlstra, A. A., Oudmaijer, R. D., et al. 2011, A\&A,534, L10

\bibitem[2003]{lobel}
Lobel, A., Dupree, A. K., Stefanik, R. P., et al.  2003, ApJ, 583, 923

\bibitem[2002]{machado}
Machado, M. A. D., de Ara\'{u}jo, F. X., Pereira, C. B. \& Fernandes, M. 
B. 2002, A\&A, 387, 151

\bibitem[2011]{maund}
Maund, J. R., Fraser, M. Ergon, M., et al. 2011, ApJ, 739, L37

\bibitem[2013]{mehner}
Mehner, A., Baade, D., Rivinius, T. et al. 2013, A\&A, 555, A116

\bibitem[2005]{messineo}
Messineo, M., Habing, H. J., Menten, K. M. et al. 2005, A\&A, 435, 575

\bibitem[2011]{millour}
Millour, F., Meilland, A., Chesneau, O., et al. 2011, A\&A, 526, A107

\bibitem[1997]{munari97}
Munari, U. \& Tomasella, L. 1997 A\&AS, 137, 521

\bibitem[2009]{munari}
Munari, U., Siviero, A., Bienaym\'{e}, O., et al. 2009, A\&A, 503, 511

\bibitem[2010]{negueruela10}
Negueruela, I., Gonz\'{a}lez-Fern\'{a}ndez, C., Marco, A., Clark, J. S., Mart\'{i}nez-N\'{u}\~{n}ez, S. 
2010, A\&A, 513, A74

\bibitem[2011]{negueruela11}
Negueruela, I., Gonz\'{a}lez-Fern\'{a}ndez, C., Marco, A. \& Clark, J. S.
2011, A\&A, 528, A59

\bibitem[2012]{negueruela12}
Negueruela, I., Marco, A., Gonz\'{a}lez-Fern\'{a}ndez, C. et al. 
2012, A\&A, 547, A15

\bibitem[2012]{nieu}
Nieuwenhuijzen, H., de Jager, C., Kolka, I., et al. 2012, A\&A, 546, A105


\bibitem[2013]{oksala}
Oksala, M. E., Kraus, M., Cidale, L. S., Muratore, M.F. \& Borges 
Fernandes, M. 2013, A\&A, 558, A17

\bibitem[2000]{okumura}
Okumura, S., Mori, A., Nishihara, E., Watanabe, E. \& Yamashita, T.
2000, ApJ, 543, 799

\bibitem[1994]{oudmaijer94}
Oudmaijer, R. D., Geballe, T. R., Waters, L. B. F. M. \& Sahu, K. C.
1994, A\&A, 281, L33

\bibitem[1996]{oudmaijer96}
Oudmaijer, R. D., Groenewegen, M. A. T., Matthews, H. E. M.,
Blommaert, J. A. D. L. \& Sahu, K. C. 1996, MNRAS, 280, 1062

\bibitem[1998]{oudmaijer98}
Oudmaijer, R. D. 1998, A\&AS, 129, 541

\bibitem[2009]{oudmaijer09}
Oudmaijer, R. D., Davies, B., de Wit, W.-J. \& Patel, M. 2009, ASPC, 412, 17

\bibitem[2013]{oudmaijer13}
Oudmaijer, R. D. \& de Wit, W.-J. 2013, A\&A, 551, A69

\bibitem[2009]{ritchie09}
Ritchie, B. W., Clark, J. S., Negueruela, I. \& Najrro, F. 2009
A\&A, 507, 1597

\bibitem[2006]{schuster}
Schuster, M. T., Humphreys, R. M. \& Marengo, M. 2006, AJ, 131, 603 

\bibitem[2010]{tiffany}
Tiffany, C., Humphreys, R. M., Jones, T. J., Davidson, K. 2010, AJ, 140, 339

\bibitem[2013]{van}
Van Dyk, S. D., Zheng W., Clubb, K. I. et al. 2013, ApJ, 772, L32

\bibitem[2010]{wachter}
Wachter, S., Mauerhan, J. C., Van Dyk, S. D. et al. 2010, AJ, 139, 2330

\bibitem[2010]{wright}
Wright, E. L., Eisenhardt, P. R. M., Mainzer, A. K. et al. 2010, AJ, 140, 1868

\bibitem[2007]{yamamuro}
Yamamuro, T., Nishimaki, Y., Motohara, K., Miyata, T. \& Tanaka, M. 2007, PASJ, 59, 
973

\end{thebibliography}
\end{document}